\documentclass[%
  reprint,
  showpacs,
  amsfonts,amssymb,amsmath,
  aps,
  pra
]{revtex4-1}

\usepackage[ascii]{inputenc}
\usepackage[T1]{fontenc}
\usepackage[english]{babel}
\usepackage{graphicx}
\usepackage{microtype}
\usepackage[pdftex,colorlinks]{hyperref}
\hypersetup{%
  linkcolor=blue,%
  citecolor=blue,%
  urlcolor=blue,%
  pdftitle={Bose-Einstein condensates with balanced gain and loss beyond
            mean-field theory},%
  pdfauthor={Dennis Dast, Daniel Haag, Holger Cartarius, J\"org Main,
             G\"unter Wunner}%
}

\DeclareMathOperator{\trace}{tr}
\DeclareMathOperator{\asin}{asin}
\DeclareMathOperator{\acos}{acos}
\newcommand{\PT}{\mathcal{PT}}
\newcommand{\mrm}{\mathrm}
\newcommand{\ha}{\hat{a}}
\newcommand{\hL}{\hat{L}}
\newcommand{\hn}{\hat{n}}
\newcommand{\hA}{\hat{A}}
\newcommand{\loss}{\mathrm{loss}}
\newcommand{\gain}{\mathrm{gain}}
\newcommand{\dt}{\frac{d}{dt}}
\newcommand{\ket}[1]{|{#1}\rangle}
\newcommand{\mean}[1]{\langle{#1}\rangle}

\begin{document}

\title{Bose-Einstein condensates with balanced gain and loss beyond mean-field
       theory}

\author{Dennis Dast}
\email[]{dennis.dast@itp1.uni-stuttgart.de}
\author{Daniel Haag}
\author{Holger Cartarius}
\author{J\"org Main}
\author{G\"unter Wunner}

\affiliation{Institut f\"ur Theoretische Physik 1,
             Universit\"at Stuttgart, 70550 Stuttgart, Germany}

\date{\today}

\begin{abstract}
  Most of the work done in the field of Bose-Einstein condensates with balanced
  gain and loss has been performed in the mean-field approximation using the
  $\PT$-symmetric Gross-Pitaevskii equation.
  In this work we study the many-particle dynamics of a two-mode condensate
  with balanced gain and loss described by a master equation in Lindblad form
  whose purity periodically drops to small values but then is nearly completely
  restored.
  This effect cannot be covered by the mean-field approximation, in which a
  completely pure condensate is assumed.
  We present analytic solutions for the dynamics in the non-interacting limit
  and use the Bogoliubov backreaction method to discuss the influence of the
  on-site interaction.
  Our main result is that the strength of the purity revivals is almost
  exclusively determined by the strength of the gain and loss and is
  independent of the amount of particles in the system and the interaction
  strength.
  For larger particle numbers, however, strong revivals are shifted towards
  longer times, but by increasing the interaction strength these strong
  revivals again occur earlier.
\end{abstract}


\maketitle

\section{Introduction}
\label{sec:introduction}

An open quantum system in which particles are injected and removed in such a
way that it nevertheless supports stationary solutions is called a quantum
system with balanced gain and loss.
Using imaginary potentials is an elegant approach to describe the in- and
outflux of particles.
If the imaginary potential is $\PT$ symmetric~\cite{Bender98a, Bender99a} or,
more generally speaking, pseudo-Hermitian~\cite{Mostafazadeh02a,
Mostafazadeh02b, Mostafazadeh02c} stationary solutions exist under certain
conditions~\cite{Fernandez98a, Fernandez14a}.

Although the concept of $\PT$ symmetry originates from quantum mechanics the
first realization of $\PT$-symmetric systems succeeded in optical
waveguides~\cite{Klaiman08a, Ruter10a, Guo09a, Peng14b}, and as a result the
focus has somewhat shifted towards optics.
However, to discuss quantum effects in systems with balanced gain and loss it
is necessary to study a genuine quantum system.

A promising candidate for the realization of a genuine quantum system with
balanced gain and loss is a Bose-Einstein condensate in a double-well potential
with an influx of particles in one well and an outflux from the other.
Such a system can be described in the mean-field limit by the Gross-Pitaevskii
equation, where balanced gain and loss is introduced through a $\PT$-symmetric
imaginary potential.
This has been investigated using a double-$\delta$
potential~\cite{Cartarius12b} and a spatially extended double
well~\cite{Dast13a}, and in both cases stationary stable solutions were found.
Furthermore proposals for the realization of such a system exist by embedding
the $\PT$-symmetric double well in a larger Hermitian
system~\cite{Kreibich13a}.

These studies were performed in the mean-field limit and thus it was assumed
that all particles are in the condensed phase, i.e., the single-particle
density matrix is quantum mechanically pure.
However, the purity of a condensate is reduced by both the coupling to the
environment and the interaction of the particles~\cite{Ruostekoski98a}.
Furthermore in systems with balanced gain and loss we are especially interested
in an exchange of particles with the environment, thus, it cannot be expected
that a description in the mean-field limit is appropriate.
To check this expectation it is necessary to carry out an analysis in the
many-particle system.

A possible many-particle description is given by a Bose-Hubbard model with
$\PT$-symmetric complex on-site energies~\cite{Graefe08b, Graefe12a}, however,
the mean-field limit of this approach does not yield the Gross-Pitaevskii
equation but only a similar equation, in which the nonlinear term is divided by
the norm squared of the state.
Although the normalized stationary solutions are the same as for the
Gross-Pitaevskii equation the dynamics differ~\cite{Dast13b, Haag14a} and,
thus, it is not the many-particle description we are looking for.

Introducing the gain and loss terms via a master equation in Lindblad
form~\cite{Breuer02a} where the coherent dynamics is governed by the
Bose-Hubbard Hamiltonian does yield the Gross-Pitaevskii equation with
imaginary potentials in the mean-field limit~\cite{Trimborn08a, Witthaut11a,
Dast14a}.
The strengths of the imaginary potentials are then given by the rate of the
Lindblad superoperators.
Master equations are routinely used to describe phase noise and both feeding
and depleting of a Bose-Einstein condensate~\cite{Anglin97a, Ruostekoski98a}.

By choosing the rate of the Lindblad superoperators in an appropriate manner we
obtain a quantum master equation with balanced gain and loss which has been
introduced in~\cite{Dast14a}.
It was shown that characteristic properties of $\PT$-symmetric systems such as
the in-phase pulsing between the lattice sites are also supported by the master
equation.
Comparing the time evolution of expectation values such as the particle number
showed that there is an excellent agreement between the results of the master
equation with balanced gain and loss and the $\PT$-symmetric Gross-Pitaevskii
equation.

However, the many-particle dynamics reveals that the condensate can differ
substantially from a completely pure condensate as assumed in the mean-field
approximation, which we could show in~\cite{Dast16a}.
In fact a periodic revival of the condensate's purity is observed where the
purity drops to small values but afterwards is nearly completely restored.
These oscillations were found to be in phase with the oscillations of the total
particle number.
These results are relevant, for instance, in the context of continuous atom
lasers if the pumping and outcoupling occurs at different
sites~\cite{Robins08a}.

Such a collapse and revival of a condensate has already been observed in an
optical lattice after ramping up the potential barrier to inhibit
tunneling~\cite{Greiner02a}.
However, these revivals occur due to the interaction between the particles and
are damped by particle losses~\cite{Pawlowski10a, Sinatra98a}, which stands in
contrast to the purity oscillations found in systems with balanced gain and
loss.
In the latter case the coupling to the environment is the driver behind the
revivals~\cite{Dast16a}.
Also it was shown that the purity of a Bose-Einstein condensate with
dissipation and phase noise can show a single revival before it decays
afterwards~\cite{Witthaut08a, Witthaut09a, Witthaut11a}.

In this paper we deepen the discussion of purity oscillations of quantum
systems with balanced gain and loss by using the Bogoliubov backreaction
method~\cite{Vardi01a, Anglin01a} in addition to directly calculating the
many-particle dynamics with the master equation.
The Bogoliubov backreaction method yields a closed set of differential
equations for the elements of the single-particle density matrix and the
covariances.
Without interaction between the particles the dynamics can be solved
analytically, and with interaction it is numerically much less costly.
We will see that for a limited time span the Bogoliubov backreaction method is
in excellent agreement with the results of the master equation and thus allows
us to extend the discussion to parameter regimes that are numerically not
accessible using the master equation.

The paper is ordered as follows.
In Sec.~\ref{sec:theory} the master equation with balanced gain and loss is
introduced and the Bogoliubov backreaction method is extended to systems with
gain and loss.
The dynamics of the non-interacting limit is solved analytically in
Sec.~\ref{sec:linear}, which allows us to understand many effects also present
with interaction.
In Sec.~\ref{sec:accuracy} the accuracy and the limitations of the Bogoliubov
backreaction method are discussed by comparison with the results of the master
equation.
A detailed study of the purity revivals for different initial states and the
influence of the initial particle number and the interaction strength follows
in Sec.~\ref{sec:revivals}.
Finally, calculating the eigenvector to the macroscopic eigenvalue of the
single-particle density matrix in Sec.~\ref{sec:eigenvectors} allows a direct
comparison with the mean-field state.
Conclusions are drawn in Sec.~\ref{sec:conclusion}.

\section{Two-mode system with balanced gain and loss}
\label{sec:theory}

The many-particle description of a Bose-Einstein condensate with balanced gain
and loss introduced in~\cite{Dast14a} is given by a quantum master equation in
Lindblad form.
It describes a system consisting of two lattice sites with loss at site 1 and
gain at site 2.
The master equation is given by
\begin{subequations}
  \begin{gather}
    \dt\hat\rho = -i [\hat H,\hat\rho]
    + \mathcal{L}_\loss\hat\rho + \mathcal{L}_\gain\hat\rho,
    \label{eq:mastereq}\\
    \hat H = -J(\ha_1^\dagger \ha_2 + \ha_2^\dagger \ha_1)
    + \frac{U}{2} (\ha_1^\dagger \ha_1^\dagger \ha_1 \ha_1
    + \ha_2^\dagger \ha_2^\dagger \ha_2 \ha_2),
    \label{eq:bhhamiltonian}\\
    \mathcal{L}_\loss \hat\rho = -\frac{\gamma_\loss}{2}
    (\ha_1^\dagger \ha_1 \hat\rho + \hat\rho \ha_1^\dagger \ha_1
    - 2 \ha_1 \hat\rho \ha_1^\dagger),
    \label{eq:liouvillianloss}\\
    \mathcal{L}_\gain \hat\rho = -\frac{\gamma_\gain}{2}
    (\ha_2 \ha_2^\dagger \hat\rho + \hat\rho \ha_2 \ha_2^\dagger
    - 2 \ha_2^\dagger \hat\rho \ha_2),
    \label{eq:liouvilliangain}
  \end{gather}
  \label{eq:completemaster}%
\end{subequations}
where the bosonic creation and annihilation operators $\ha_j^\dagger$ and
$\ha_j$ are used and we set $\hbar = 1$.

The coherent dynamics of bosonic atoms in the lowest-energy Bloch band of an
optical lattice are described by the Bose-Hubbard Hamiltonian~\cite{Jaksch98a}
in Eq.~\eqref{eq:bhhamiltonian}.
In the two-mode formulation the Hamiltonian is well suited to describe a
Bose-Einstein condensate in a double-well potential~\cite{Gati07a}.
The first term of Eq.~\eqref{eq:bhhamiltonian} describes tunneling between the
two lattice sites and the second term describes an on-site interaction between
the particles.
The strength of the tunneling is given by the parameter $J$, which for all
results shown in this work is taken to be $J=1$.
This choice effectively sets the time scale to $\tau = \hbar/J$.
To define the strength of the on-site interaction we use the macroscopic
interaction strength
\begin{equation}
  g=(N_0-1)U,
  \label{eq:macroscopicinteraction}
\end{equation}
which depends on the initial amount of particles $N_0$ in the system.

The controlled outcoupling of particles at lattice site 1 could be realized by
a focused electron beam.
It was demonstrated that using a commercial electron microscope it is possible
to remove atoms from single sites of an optical lattice~\cite{Gericke08a,
Wurtz09a, Barontini13a}.
If an incident electron collides with an atom, the atom is ionized or excited
and escapes the trapping potential.
Additionally there are secondary collisions which lead to further atom losses.
Since it was shown that there is almost no heating due to the electron beam it
can be seen as an almost pure dissipative effect describable by a Lindblad
superoperator of the form~\eqref{eq:liouvillianloss}.

A continuous and coherent incoupling of atoms into a Bose-Einstein condensate
was experimentally realized by feeding from a second
condensate~\cite{Robins08a}, thus being a possible realization for the particle
gain at lattice site 2.
In this setup the second condensate acts as a source of particles and is
located above the first condensate.
By applying a continuous radiofrequency field, atoms in the source condensate
make a transition from an $m_\mrm{F} = 2$ state to an $m_\mrm{F} = 0$ state.
As a result they leave the magneto-optical trap and begin to fall under the
action of gravity towards the lower condensate.
An upward propagating light beam causes a transition of the falling atoms into
a state from which they are stimulated to emit into the state of the lower
condensate.
A subsequent study indicated that the pumping occurs in a Raman
superradiance-like process~\cite{Doring09a, Schneble04a, Yoshikawa04a}.

The parameters $\gamma_\loss$ and $\gamma_\gain$ determine the strength of the
particle gain and loss.
They are balanced in the following way
\begin{equation}
  \gamma_\loss = \frac{N_0+2}{N_0}\gamma_\gain \equiv \gamma,
  \label{eq:gainlossratio}
\end{equation}
which ensures that if half of the particles are at each lattice site then, for
short times, the particle gain and loss have equal strength~\cite{Dast14a}.
In this work we will use the abbreviation $\gamma \equiv \gamma_\loss$ to
define the strength of the in- and outcoupling and $\gamma_\gain$ is always
chosen such that it fulfills Eq.~\eqref{eq:gainlossratio}.

The Bogoliubov backreaction method~\cite{Vardi01a, Anglin01a} allows us to
calculate the time evolution of the single-particle density matrix instead of
solving the many-particle dynamics.
The time derivative of the single-particle density matrix $\sigma_{j k} =
\mean{\ha_j^\dagger \ha_k}$ calculated with the master
equation~\eqref{eq:completemaster} reads
\begin{align}
  \dt\sigma_{j k} =
  &-i J (\sigma_{j+1\,k} + \sigma_{j-1\,k}- \sigma_{j\,k+1} - \sigma_{j\,k-1})
  \notag\\
  &- i U (\sigma_{k k}\sigma_{j k} - \sigma_{j j}\sigma_{j k}
  + \Delta_{j k k k} - \Delta_{j j j k})
  \notag\\
  &- \frac{\gamma_{\loss,j} + \gamma_{\loss,k}}{2} \sigma_{j k}
  \notag\\
  &+ \frac{\gamma_{\gain,j} + \gamma_{\gain,k}}{2} (\sigma_{j k} +
  \delta_{j k}),
  \label{eq:dspdm}
\end{align}
where $\gamma_{\gain,j}$ and $\gamma_{\loss,j}$ are the strength of the gain
and loss contributions at lattice site $j$.
In this representation Eq.~\eqref{eq:dspdm} is more general than
Eqs.~\eqref{eq:completemaster} since it holds for a Bose-Hubbard chain with
arbitrary length and gain and loss at arbitrary lattice sites.

However, Eq.~\eqref{eq:dspdm} does not yield a closed set of differential
equations because of the covariances
\begin{equation}
  \Delta_{j k l m} =
  \mean{\ha_j^\dagger \ha_k \ha_l^\dagger \ha_m}
  - \mean{\ha_j^\dagger \ha_k} \mean{\ha_l^\dagger \ha_m}.
  \label{eq:covariances}
\end{equation}
Neglecting the covariances, i.e.\ approximating second-order expectation values
by a product of first-order expectation values $\mean{\ha_j^\dagger \ha_k
\ha_l^\dagger \ha_m} \approx \mean{\ha_j^\dagger \ha_k} \mean{\ha_l^\dagger
\ha_m}$, would lead to a closed set of differential equations but is only valid
for large particle numbers and close to pure condensates.

The Bogoliubov backreaction improves the approximation by taking the time
evolution of the covariances into account.
This method has been successfully used for closed systems~\cite{Vardi01a,
Anglin01a, Tikhonenkov07a} and systems with dissipation~\cite{Witthaut11a},
thus, we expect to obtain also accurate results for systems with both gain and
loss.
Calculating the time derivative of the covariances will again result in
expectation values of higher order due to the nonlinear term.
These third-order expectation values are approximated by a product of
first-order and second-order expectation values in the following
way~\cite{Vardi01a}
\begin{align}
  & \mean{\ha_i^\dagger \ha_j \ha_k^\dagger \ha_l
  \ha_m^\dagger \ha_n} \approx
  \mean{\ha_i^\dagger \ha_j \ha_k^\dagger
  \ha_l}\mean{\ha_m^\dagger \ha_n}
  \notag\\
  &\quad + \mean{\ha_i^\dagger \ha_j \ha_m^\dagger
  \ha_n}\mean{\ha_k^\dagger \ha_l}
  + \mean{\ha_k^\dagger \ha_l \ha_m^\dagger
  \ha_n}\mean{\ha_i^\dagger \ha_j}
  \notag\\
  &\quad - 2 \mean{\ha_i^\dagger \ha_j}
  \mean{\ha_k^\dagger \ha_l} \mean{\ha_m^\dagger \ha_n}.
  \label{eq:thirdordermoments}
\end{align}
Using the approximation~\eqref{eq:thirdordermoments} the time derivatives of
the covariances read
\begin{align}
  & \dt \Delta_{j k l m} =
  \notag\\
  &\quad - i J (\Delta_{j+1\,k l m} + \Delta_{j-1\,k l m}
  - \Delta_{j\,k+1\,l m} - \Delta_{j\,k-1\,l m}
  \notag\\
  &\qquad + \Delta_{j k\,l+1\,m} + \Delta_{j k\,l-1\,m}
  - \Delta_{j k l\,m+1} - \Delta_{j k l\,m-1})
  \notag\\
  &\quad + i U [\Delta_{j k l m} (\sigma_{j j} - \sigma_{k k} + \sigma_{l l}
  - \sigma_{m m})
  \notag\\
  &\qquad + \Delta_{j j l m}\sigma_{j k} - \Delta_{k k l m}\sigma_{j k}
  + \Delta_{j k l l}\sigma_{l m} - \Delta_{j k m m}\sigma_{l m}]
  \notag\\
  &\quad - \frac{\gamma_{\loss,j} + \gamma_{\loss,k} +
  \gamma_{\loss,l} + \gamma_{\loss,m}}{2} \Delta_{j k l m}
  \notag\\
  &\qquad + \delta_{k l} \gamma_{\loss,k} \sigma_{j m}
  \notag\\
  &\quad + \frac{\gamma_{\gain,j} + \gamma_{\gain,k} +
  \gamma_{\gain,l} + \gamma_{\gain,m}}{2} \Delta_{j k l m}
  \notag\\
  &\qquad + \delta_{j m} \gamma_{\gain,j} (\sigma_{l k} + \delta_{l k}),
  \label{eq:dcov}
\end{align}
which form, together with Eq.~\eqref{eq:dspdm}, a closed set of differential
equations.

For $M$ lattice sites Eq.~\eqref{eq:dspdm} yields $M^2$ complex equations and
Eq.~\eqref{eq:dcov} $M^4$ complex equations.
However, the single-particle density matrix contains only $M^2$ independent
real elements due to $\sigma_{j k} = \sigma_{k j}^*$, and there are only
$\frac12 (M^4+M^2)$ independent real quantities for the covariances due to
$\Delta_{j k l m} = \Delta_{m l k j}^*$ and $\Delta_{j k l m}
= \Delta_{l m j k} - \delta_{j m}\sigma_{l k} + \delta_{l k}\sigma_{j m}$.
When choosing the independent covariances one has to keep in mind that
commutation relations for the indices of the covariances do not necessarily
hold for their time derivatives given by Eq.~\eqref{eq:dcov} as a result of
the approximation~\eqref{eq:thirdordermoments}.

For the two-mode system considered here we can use the Bloch
representation~\cite{Anglin01a}.
The four real independent quantities of the single-particle density matrix are
then given by the expectation values $s_j = 2\mean{\hL_j}$ and $n = \mean{\hn}$
of the four Hermitian operators,
\begin{subequations}
  \begin{align}
    \hL_x &= \frac{1}{2} (\ha_1^\dagger \ha_2 + \ha_2^\dagger \ha_1),&
    \hL_y &= \frac{i}{2} (\ha_1^\dagger \ha_2 - \ha_2^\dagger \ha_1),\\
    \hL_z &= \frac{1}{2} (\ha_2^\dagger \ha_2 - \ha_1^\dagger \ha_1),&
    \hn &= \ha_1^\dagger \ha_1 + \ha_2^\dagger \ha_2,
  \end{align}
  \label{eq:blochoperators}%
\end{subequations}
and the ten real independent covariances are the covariances between these
operators
\begin{equation}
  \Delta_{j k} = \mean{\hA_j \hA_k + \hA_k \hA_j} - 2 \mean{\hA_j}\mean{\hA_k},
  \label{eq:blochcov}
\end{equation}
where $\hA_{j} \in \{\hL_x, \hL_y, \hL_z, \hn\}$.
The time derivatives of $s_j$ and $\Delta_{j k}$ are then given by linear
combinations of Eqs.~\eqref{eq:dspdm} and \eqref{eq:dcov}, respectively.
They read
\begin{subequations}
  \begin{align}
    \dot{s}_x &= -U (s_y s_z + 2\Delta_{yz}) - \gamma_- s_x, \\
    \dot{s}_y &= 2 J s_z + U(s_x s_z + 2\Delta_{xz}) - \gamma_- s_y, \\
    \dot{s}_z &= -2 J s_y + \gamma_+ n - \gamma_- s_z + \gamma_\gain, \\
    \dot{n}   &= -\gamma_- n + \gamma_+ s_z + \gamma_\gain,
    \label{eq:dspdmblochn}
  \end{align}
  \label{eq:dspdmbloch}%
\end{subequations}
\begingroup
\allowdisplaybreaks
\begin{subequations}
  \begin{align}
    \dot{\Delta}_{xx} =& -2 U (s_z \Delta_{xy} + s_y \Delta_{xz}) \notag\\
                       & - \gamma_- (2\Delta_{xx} - \frac{s_z}{2})
                         + \gamma_+ \frac{n}{2} + \frac{\gamma_\gain}{2}, \\
    \dot{\Delta}_{yy} =& + 4 J \Delta_{yz}
                         + 2 U (s_z \Delta_{xy} + s_x \Delta_{yz}) \notag\\
                       & - \gamma_- (2\Delta_{yy} - \frac{s_z}{2})
                         + \gamma_+ \frac{n}{2} + \frac{\gamma_\gain}{2}, \\
    \dot{\Delta}_{zz} =& -4 J \Delta_{yz}
                         - \gamma_-(2\Delta_{zz} + \frac{s_z}{2}) \notag\\
                       & + \gamma_+(\Delta_{zn} + \frac{n}{2})
                         + \frac{\gamma_\gain}{2}, \\
    \dot{\Delta}_{xy} =& + 2 J \Delta_{xz}
                         + U(s_x \Delta_{xz} + s_z \Delta_{xx} \notag\\
                       & - s_z \Delta_{yy} - s_y \Delta_{yz})
                         - 2 \gamma_- \Delta_{xy}, \\
    \dot{\Delta}_{xz} =& - 2 J \Delta_{xy}
                         - U(s_y \Delta_{zz} + s_z \Delta_{yz}) \notag\\
                       & - \gamma_- (2\Delta_{xz} + \frac{s_x}{2})
                         + \gamma_+ \frac{\Delta_{xn}}{2}, \\
    \dot{\Delta}_{yz} =& + 2 J (\Delta_{zz} - \Delta_{yy})
                         + U (s_x \Delta_{zz} + s_z \Delta_{xz}) \notag\\
                       & - \gamma_- (2\Delta_{yz} + \frac{s_y}{2})
                         + \gamma_+ \frac{\Delta_{yn}}{2}, \\
    \dot{\Delta}_{xn} =& -U (s_z \Delta_{yn} + s_y \Delta_{zn}) \notag\\
                       & - 2 \gamma_- \Delta_{xn}
                         + \gamma_+ (2 \Delta_{xz} + s_x),
                       \label{eq:dcovblochxn}\\
    \dot{\Delta}_{yn} =& + 2 J \Delta_{zn}
                         + U (s_x \Delta_{zn} + s_z \Delta_{xn}) \notag\\
                       & - 2 \gamma_- \Delta_{yn}
                         + \gamma_+ (2\Delta_{yz} + s_y), \\
    \dot{\Delta}_{zn} =& - 2 J \Delta_{yn} - \gamma_- (2 \Delta_{zn} + n)
                         \notag\\
                       & + \gamma_+ (2 \Delta_{zz} + \frac{\Delta_{nn}}{2}
                         + s_z) + \gamma_\gain, \\
    \dot{\Delta}_{nn} =& -\gamma_- (2 \Delta_{nn} + 2 s_z)
                         + \gamma_+ (4\Delta_{zn} + 2n) \notag\\
                       & + 2 \gamma_\gain,
                       \label{eq:dcovblochnn}
  \end{align}
  \label{eq:dcovbloch}%
\end{subequations}
\endgroup
with $\gamma_- = (\gamma_\loss-\gamma_\gain)/2$ and $\gamma_+ =
(\gamma_\loss+\gamma_\gain)/2$.
The differential equations of the first-order moments~\eqref{eq:dspdmbloch} and
the second-order moments~\eqref{eq:dcovbloch} are coupled via the nonlinear
interaction term, i.e., terms containing the parameter $U$.
Furthermore the differential equation of $n$~\eqref{eq:dspdmblochn} and the
covariances of $n$~\eqref{eq:dcovblochxn}--\eqref{eq:dcovblochnn} are
coupled to the remaining equations only by terms containing $\gamma$, which is
not surprising since these terms arise due to the gain and loss of particles,
and without them the particle number is conserved.
The inhomogeneities of all differential equations contain only $\gamma_\gain$,
thus, they solely arise due to the particle gain.

Starting from Eq.~\eqref{eq:dspdm} the mean-field approximation is obtained by
assuming a pure condensate, i.e.\ $\sigma_{j k} = c_j^* c_k$ with the
mean-field occupation coefficients $c_i$ of site $i$, and neglecting the
covariances.
Both assumptions hold in the limit $N_0 \to \infty$.
For the two-mode system with balanced gain and loss~\eqref{eq:completemaster}
the mean-field limit is the $\PT$-symmetric Gross-Pitaevskii
equation~\cite{Dast14a},
\begin{subequations}
  \begin{align}
    i \dt c_1 &= - J c_2 + g |c_1|^2 c_1
      - i \frac{\gamma}{2} c_1,\\
    i \dt c_2 &= - J c_1 + g |c_2|^2 c_2
      + i \frac{\gamma}{2} c_2.
  \end{align}
  \label{eq:discreteGPE}%
\end{subequations}
In the mean-field limit a state is defined by two complex numbers $c = (c_1,\
c_2)^T$ whereat only two real degrees of freedom remain due to normalization
and the choice of a global phase.
The corresponding many-particle state with a total number of $N_0$ particles
is, in the Fock basis $\ket{n_1, n_2}$, given by~\cite{Dast14a}
\begin{equation}
  \ket{\psi} = \sum_{m=0}^{N_0}
  \sqrt{ \begin{pmatrix} N_0 \\ m \end{pmatrix} }
  c_1^{N_0-m}c_2^m \ket{N_0-m,m}.
  \label{eq:productState}
\end{equation}

The Gross-Pitaevskii equation~\eqref{eq:discreteGPE} supports two
$\PT$-symmetric stationary solutions
\begin{equation}
  c_1 = \pm \frac{1}{\sqrt{2}}\exp{\left(\pm i \asin{
  \left(\frac{\gamma}{2J}\right)}\right)}, \quad
  c_2 = \frac{1}{\sqrt{2}},
  \label{eq:stationaryStates}
\end{equation}
which exist for $|\gamma| \leq 2J$ and which we will call the ground (positive
signs) and excited state (negative signs) of the system~\cite{Graefe12a,
Dast14a}.
In addition there are two $\PT$-broken solutions for $|\gamma| \geq \sqrt{4J^2
- g^2}$.

In this work we use results obtained by directly solving the master
equation~\eqref{eq:completemaster} via the quantum jump method~\cite{Molmer93a,
Johansson13a} and results obtained by integrating the differential
equations~\eqref{eq:dspdmbloch} and \eqref{eq:dcovbloch} of the Bogoliubov
backreaction method.
For the quantum jump method an average over a certain amount of quantum
trajectories is performed till the results converge.

The essential quantity discussed in this work is the purity of the reduced
single-particle density matrix $\sigma_{\mrm{red},jk} =
\sigma_{jk}/\trace{\sigma}$, which measures how close the condensate is to a
pure Bose-Einstein condensate,
\begin{equation}
  P = 2 \trace \sigma_\mrm{red}^2 - 1 = \frac{s_x^2 + s_y^2 + s_z^2}{n^2}
  \ \in [0,1].
  \label{eq:purity}
\end{equation}
A pure condensate with $P=1$ is described by a product state as it is assumed
for the Gross-Pitaevskii equation.
In this case all particles are in the condensed mode since all but one
eigenvalue of $\sigma_\mrm{red}$ vanish.
An increasing occupation of the non-condensed mode reduces the value of $P$.
If more than one eigenvalue is large the condensate is called
fragmented~\cite{Mueller06a}.

For a given initial particle number $N_0$ a pure state has only two degrees of
freedom,
\begin{subequations}
  \begin{align}
    s_x &= N_0 \sin(\vartheta) \cos(\varphi), \\
    s_y &= N_0 \sin(\vartheta) \sin(\varphi), \\
    s_z &= N_0 \cos(\vartheta).
  \end{align}
  \label{eq:purestate}%
\end{subequations}
In the following pure states will therefore be characterized by the two angles
$\varphi$ and $\vartheta$.
They are obtained from the normalized $c_{1/2}$ by $\vartheta =
\acos(1-2|c_1|^2)$ and $\varphi = \arg(c_1 c_2^*)$.
Accordingly the two $\PT$-symmetric solutions~\eqref{eq:stationaryStates} are
given by $\vartheta = \pi/2$ and $\varphi = \pi/2 \mp \acos(\gamma/2J)$.

The purity of the condensate can be measured in interference experiments where
the double-well trap is turned off and as a result the condensate expands and
interferes~\cite{Gati06b, Shin04a}.
The average contrast~\cite{Gati06b, Witthaut08a, Witthaut09a} in such an
experiment is obtained by averaging over various realizations~\cite{Castin97a}
and is given by
\begin{equation}
  \nu = \frac{2 |\mean{\ha_1^\dagger \ha_2}|}{\mean{\ha_1^\dagger \ha_1} +
  \mean{\ha_2^\dagger \ha_2}}
  = \frac{\sqrt{s_x^2+s_y^2}}{n} \ \in [0,1].
  \label{eq:contrast}
\end{equation}
By introducing the squared imbalance of the particle number in the two lattice
sites
\begin{equation}
  I = \left( \frac{\mean{\ha_1^\dagger\ha_1} - \mean{\ha_2^\dagger\ha_2}}
  {\mean{\ha_1^\dagger\ha_1} + \mean{\ha_2^\dagger\ha_2}}
  \right)^2
  = \frac{s_z^2}{n^2} \ \in [0,1],
  \label{eq:imbalance}
\end{equation}
the squared average contrast can be written as
\begin{equation}
  \nu^2 = P - I.
  \label{eq:contrastSquared}
\end{equation}
This shows that the purity is the upper limit for the squared contrast and for
equally distributed particles the two quantities are identical.

\section{Non-interacting limit}
\label{sec:linear}

For vanishing interaction, i.e.\ $U=0$, the differential equations of the
first-order moments~\eqref{eq:dspdmbloch} and the second-order
moments~\eqref{eq:dcovbloch} decouple.
Thus, the first-order moments already yield a closed set of linear differential
equations
\begin{subequations}
  \begin{align}
    \dot{s}_x &= - \gamma_- s_x, \\
    \dot{s}_y &= 2 J s_z - \gamma_- s_y, \\
    \dot{s}_z &= -2 J s_y + \gamma_+ n - \gamma_- s_z + \gamma_\gain, \\
    \dot{n}   &= -\gamma_- n + \gamma_+ s_z + \gamma_\gain,
  \end{align}
  \label{eq:blochlinear}%
\end{subequations}
which can be solved analytically.

There is an oscillatory regime for $4J^2 > \gamma_+^2$ in which the solution
reads
\begin{subequations}
  \begin{align}
    s_x(t) &= \kappa_1 e^{-\gamma_- t},\\
    s_y(t) &= \alpha_2 + [\gamma_+ \kappa_2
              + 2 J \kappa_3 \cos(\omega t - \kappa_4)] e^{-\gamma_- t},\\
    s_z(t) &= \alpha_3 - \omega \kappa_3 \sin(\omega t - \kappa_4)
              e^{-\gamma_- t},\\
    n(t)   &= \alpha_4 + [2J \kappa_2
              + \gamma_+ \kappa_3 \cos(\omega t - \kappa_4)] e^{-\gamma_- t}
  \end{align}
  \label{eq:solosci}%
\end{subequations}
with $\omega = \sqrt{4J^2 - \gamma_+^2}$, the steady state of the system
\begin{equation}
  \boldsymbol{\alpha} =
  \frac{\gamma_+^2-\gamma_-^2}{4J^2-\gamma_+^2+\gamma_-^2}
  \begin{pmatrix}
    0 \\
    \frac{2J}{\gamma_-} \\
    1 \\
    1 + \frac{4J^2}{\gamma_-(\gamma_+ + \gamma_-)}
  \end{pmatrix},
  \label{eq:partsol}%
\end{equation}
and the four real parameters $\kappa_i$ that define the initial state.

As can be directly seen the steady state is an attractor and every trajectory
will finally reach this state with the decay rate $\gamma_-$.
For balanced gain and loss as defined in Eq.~\eqref{eq:gainlossratio} the decay
rate is given by $\gamma_- = \gamma_\gain/N_0$.

The short-term behavior is dominated by oscillations with the characteristic
frequency $\omega$.
The frequency is maximum for vanishing gain and loss, $\gamma_+ \to 0$, and
decreases to zero for $\gamma_+ \to 2J$.

Since the purity oscillations discussed in~\cite{Dast16a} are driven by the
gain and loss of the system and not by the interaction of the particles the
purity oscillations are also present in the non-interacting limit, thus giving
us access to an analytic discussion of the properties of the purity.

\begin{figure}
  \includegraphics[width=\columnwidth]{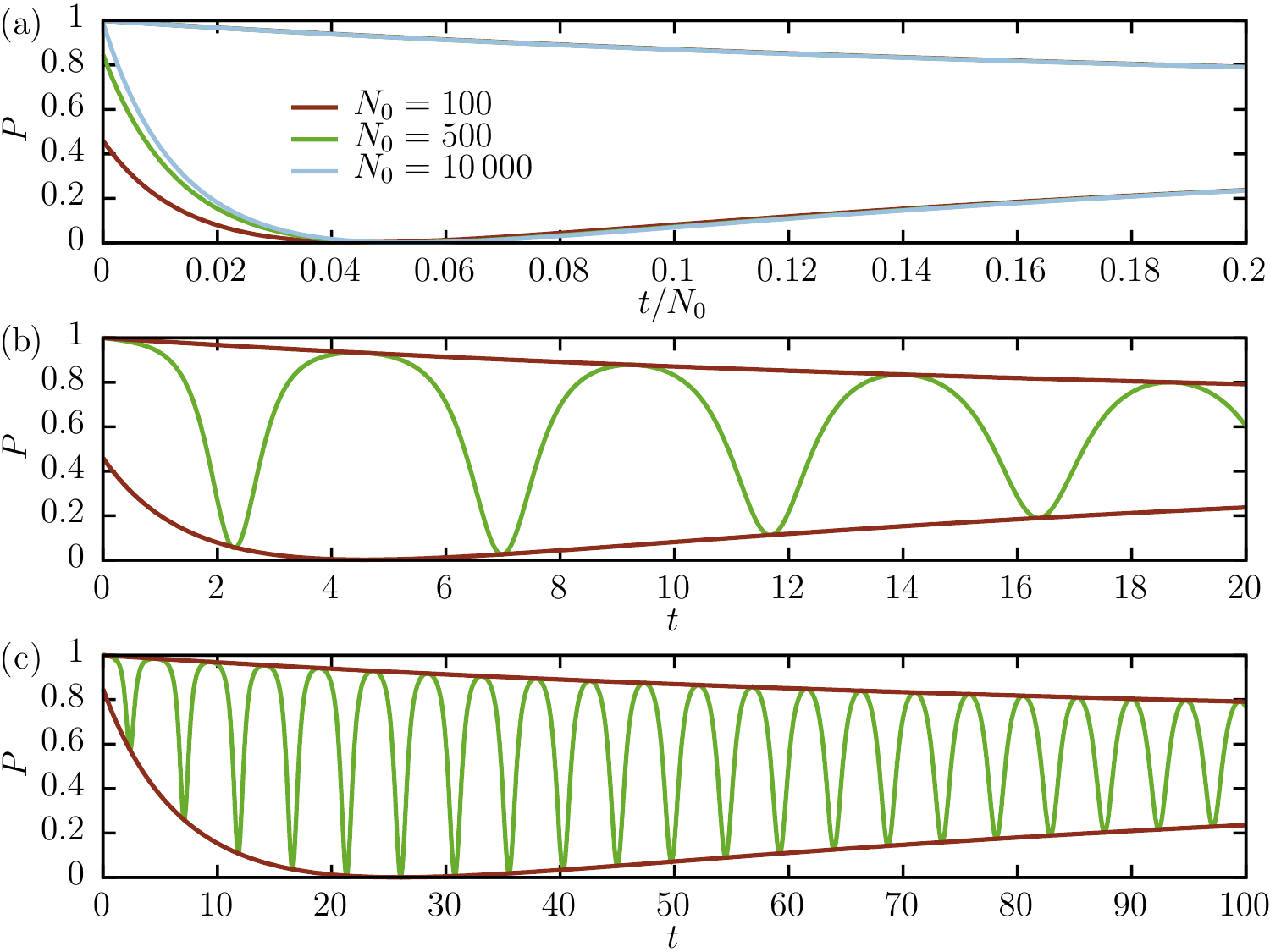}
  \caption{%
    (a) The envelope functions of the purity oscillations~\eqref{eq:envelope}
    for three different initial particle numbers in the non-interacting limit.
    The time is scaled by the particle number in such a way that the envelope
    functions become similar for large particle numbers.
    Although the particle number changes the time scale of the envelope the
    period of the actual oscillations stays approximately the same as can be
    seen for (b) $N_0=100$ and (c) $N_0=500$.
    In all calculations the pure initial state
    $\varphi = \vartheta = \pi/2$ and the gain-loss parameter $\gamma=1.5$
    were used.
  }
  \label{fig:purityLinear}
\end{figure}

The dynamical behavior of the purity consists of fast oscillations with
frequency $\omega$, which are confined by an envelope function as shown in
Figs.~\ref{fig:purityLinear}(b) and (c).
The lower and upper envelope functions are very precisely given by
\begin{equation}
  P_\mrm{l/u} = \frac{\kappa_1^2 +
    (\alpha_2 e^{\gamma_- t} + \gamma_+ \kappa_2 \mp 2J |\kappa_3|)^2
    + \alpha_3^2 e^{2\gamma_- t}}
    {(\alpha_4 e^{\gamma_- t} + 2J\kappa_2
    \mp \gamma_+ |\kappa_3|)^2}.
  \label{eq:envelope}%
\end{equation}
These envelope functions are obtained by calculating the
purity~\eqref{eq:purity} for the solutions~\eqref{eq:solosci} and setting
$\kappa_3 \cos(\omega t - \kappa_4) = \mp |\kappa_3|$ and
$\kappa_3 \sin(\omega t - \kappa_4) = 0$.

Since the initial particle number $N_0$ is much larger than the system
parameters $J$ and $\gamma$, and the initial values for $s_{x,y,z}$ all scale
with $N_0$, we can expand Eq.~\eqref{eq:envelope} in powers of $N_0$.
This is done for all terms except the time dependent term $e^{\gamma_- t}$
since $t$ might be large.
The calculation shows that the leading order of both the numerator and the
denominator is $N_0^2$.
By neglecting all other orders the remaining influence of $N_0$ is only in the
exponential term $e^{\gamma_- t}$.
With Eq.~\eqref{eq:gainlossratio} we can write $\gamma_- =
\gamma_\loss/(N_0+2) \approx \gamma_\loss/N_0$ and thus
$e^{\gamma_- t} \approx e^{\gamma_\loss t/N_0}$.

This shows that for $N_0 \gg 1$ the initial particle number only changes the
time scale of $P_\mrm{l/u}$.
Multiplying $N_0$ with a factor effectively stretches the time scale by this
factor.
Thus, the dynamics of the envelope functions are slower for higher particle
numbers.
Since the envelope functions define the strength of the purity revivals we can
conclude that not the strength of the revivals is changed by $N_0$ but only the
time at which strong revivals occur.

This can be checked in Fig.~\ref{fig:purityLinear}(a), which shows the envelope
functions for three different initial particle numbers $N_0$.
The rescaled time parameter $t/N_0$ is used so that we expect all envelopes
to become similar for $N_0 \gg 1$.
In fact the upper envelope function is virtually identical for all particle
numbers.
The lower envelope function is different in the initial area and especially
$P_\mrm{l}(t=0)$ has different values.
This difference, however, vanishes for large particle numbers and consequently
the difference between $N_0=10\,000$ and $N_0=500$ is much smaller than that
between $N_0=500$ and $N_0=100$.
For $t/N_0 \gtrsim 0.04$ also the lower envelope functions lie almost perfectly
on top of each other.

Note that the frequency of the fast oscillations with $\omega = \sqrt{4 J^2 -
\gamma_+^2}$ is mostly unaffected by $N_0$ since $\gamma_+ = \gamma_\loss
(N_0+1)/(N_0+2) \approx \gamma_\loss$.
This explains the behavior shown in Figs.~\ref{fig:purityLinear}(b) and (c),
which compares the purity oscillations for $N_0=100$ and $N_0=500$.
The oscillation frequency is approximately the same but the time scale of the
envelope functions in (c) is stretched by a factor of $5$.
As a result the first purity revivals for $N_0=500$ are small and they become
only stronger once the difference of the envelope functions becomes larger,
whereas for $N_0=100$ already the first purity revival is strong.

\section{Accuracy of the Bogoliubov backreaction method}
\label{sec:accuracy}

With interaction between the particles the differential equations of first
order couple to the second-order moments and we can no longer solve this
set of nonlinear differential equations analytically.
Instead Eqs.~\eqref{eq:dspdmbloch} and \eqref{eq:dcovbloch} are integrated
numerically.
It is not a priori clear that the Bogoliubov backreaction method yields precise
results for this system since the expansion in higher order expectation values
converges in powers of the smaller eigenvalue of
$\sigma_\mrm{red}$~\cite{Vardi01a, Anglin01a}.
Thus we can only be sure to obtain accurate results as long as the condensate
remains almost pure, i.e.\ the matrix $\sigma_\mrm{red}$ has one eigenvalue
close to one and the remaining eigenvalue is close to zero.

\begin{figure}
  \includegraphics[width=\columnwidth]{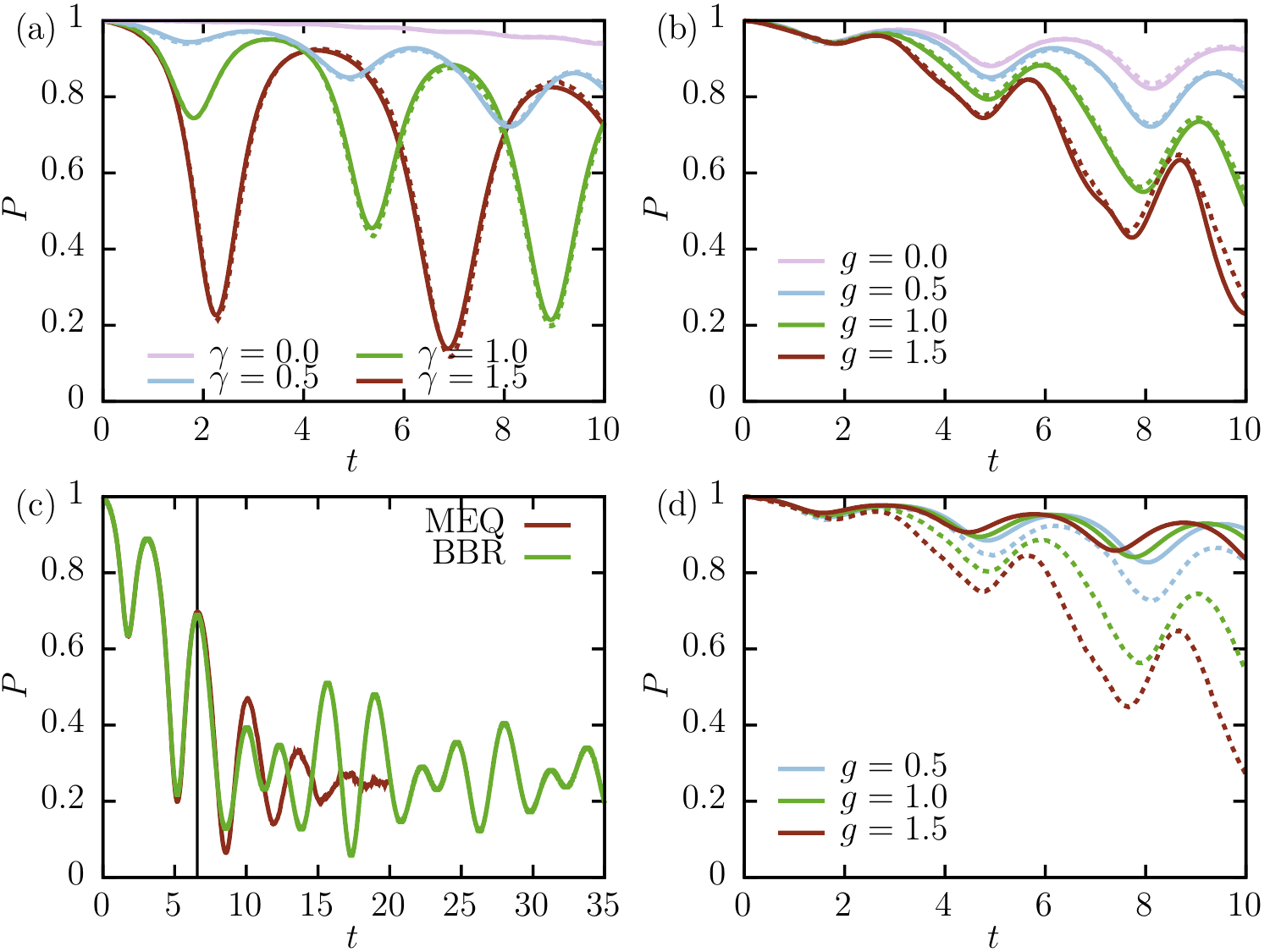}
  \caption{%
    Time evolution of the purity $P$ for different values of (a) the
    gain-loss parameter $\gamma$ (with $g=0.5$) and (b) the interaction
    strength $g$ (with $\gamma=0.5$).
    The results obtained with the Bogoliubov backreaction method (solid lines)
    are in excellent agreement with the results obtained with the many-particle
    calculations (dotted lines) but for stronger interaction the agreement is
    slightly worse.
    In these calculations the pure initial state
    $\varphi=\vartheta=\pi/2$ and the particle number $N_0=100$ were
    used.
    For the master equation it was averaged over 500 trajectories.
    For longer times (c) the Bogoliubov backreaction method (BBR) shows a
    behavior similar to a beat frequency that is not observed using the master
    equation (MEQ).
    This is used to define the last maximum of the purity where the revival
    strength still increases as the limit for the reliability of the Bogoliubov
    backreaction method (marked by the black vertical line).
    The parameters used are $\varphi=\vartheta=\pi/2$, $\gamma=1$,
    $g=1$, $N_0=50$ and for the master equation it was averaged over 3000
    trajectories.
    (d) Neglecting the covariances in Eq.~\eqref{eq:dspdm} (solid lines) shows
    a behavior that differs substantially from the many-particle calculations
    (dotted lines).
    The parameters used are $\gamma = 0.5$ and $N_0 = 100$.
  }
  \label{fig:bbrTest}
\end{figure}

To evaluate the accuracy of the Bogoliubov backreaction method it is compared
with results directly obtained using the master
equation~\eqref{eq:completemaster}.
Figure~\ref{fig:bbrTest}(a) compares the results of the two approaches for
different values of the gain-loss parameter with constant interaction strength
$g=0.5$.
The results obtained with the Bogoliubov backreaction method (dotted lines) lie
nearly perfectly on top of the results obtained with the master equation (solid
lines) for all trajectories shown.
Comparing all elements of the single-particle density matrix and the
covariances confirms this excellent agreement.
It is especially remarkable that even for $\gamma=1.5$ where the purity of the
condensate drops to very small values, thus, violating the aforementioned
condition, the numerical results nevertheless show this excellent agreement.

Since the approximation of the Bogoliubov backreaction method is applied to the
nonlinear interaction term it is expected that the approximation becomes worse
for stronger interactions.
Figure~\ref{fig:bbrTest}(b) shows the time evolution of the purity for
different values of $g$ but with an identical gain-loss parameter $\gamma =
0.5$ and compares the results of the master equation with those of the
Bogoliubov backreaction method.
Note that there is even a small discrepancy for $g=0$ where the dynamics of the
single-particle density matrix~\eqref{eq:dspdmbloch} is exact.
This discrepancy stems from the fact that the quantum jump method is not exact
itself but only becomes exact in the limit of infinitely many quantum
trajectories.
As expected for stronger nonlinearities the discrepancy between the two
different approaches becomes slightly larger but is still very good.

There is, however, a fundamental difference between the results of the two
approaches for longer times as illustrated by a sample trajectory in
Fig.~\ref{fig:bbrTest}(c).
After a few purity oscillations the Bogoliubov backreaction method shows a
behavior similar to a beat frequency where the amplitude of the oscillations
increases and decreases periodically.
This behavior is not found using the many-particle calculations at all.
We can now use this observation to identify a limit for the reliability of the
Bogoliubov backreaction method.
For all trajectories checked we found that the purity revival from one minimum
to the subsequent maximum increases for the first oscillations, i.e.\ $\Delta
P_i = P_{\mrm{max},i} - P_{\mrm{min},i}$ increases with $i$.
It decreases for the first time when the first node of the beat frequency is
approached.
Thus, we use the last maximum where the revival strength still increases as the
limit for the Bogoliubov backreaction method.
This limit is visualized in Fig.~\ref{fig:bbrTest}(c) by the vertical black
line.

Since within this limit there is an excellent agreement between the Bogoliubov
backreaction method and the many-particle dynamics one might ask if it is even
necessary to take the covariances into account.
As mentioned in Sec.~\ref{sec:theory} a closed set of equations for the
single-particle density matrix is also obtained if we neglect the covariances
in Eq.~\eqref{eq:dspdmbloch}, which is equivalent to the approximation
$\mean{\ha_j^\dagger \ha_k \ha_l^\dagger \ha_m} \approx \mean{\ha_j^\dagger
\ha_k} \mean{\ha_l^\dagger \ha_m}$.

The results obtained in this approximation are shown in
Fig.~\ref{fig:bbrTest}(d) (solid lines) compared with the many-particle
dynamics (dotted lines) for different values of the macroscopic interaction
$g$.
Using this approximation we still find oscillations of the purity and also the
frequency of the oscillation, which increases for larger values of $g$, is
well captured.
However, the actual values of the purity differ substantially.
The many-particle calculations show considerably smaller purities, and the
purities become smaller for increasing values of $g$.
If the covariances are neglected this influence of the interaction is not
found.
Instead the purity stays even closer to unity for stronger interactions.

This shows that to understand the physics of a system with balanced gain and
loss it is necessary to take the covariances into account.
The covariances are fluctuations that are driven by the single-particle density
matrix but they also yield a backreaction by the coupling terms proportional
to $U$.
These fluctuations alter the behavior of the system in an essential manner and
are the leading corrections to the dynamics since it is not necessary to
consider higher orders to obtain accurate results.

\section{Purity revivals}
\label{sec:revivals}

In this section we perform a detailed analysis of the purity revivals using
the Bogoliubov backreaction method.
Our main interest is to study how the revivals depend on the gain-loss
parameter $\gamma$ and the interaction strength $g$.
To characterize the strength of the purity oscillations we use the strongest
purity revival $\Delta P$, i.e.\ the greatest increase from a purity minimum to
the subsequent maximum.
As discussed in the previous section the Bogoliubov backreaction method yields
accurate results till the revival strength decreases, so the last purity
revival that is within reach of the Bogoliubov backreaction method can be taken
as the strongest revival.
This strongest revival, however, does not only depend on the parameters of the
system but also on the choice of the initial state.
Since the initial state is always pure, it is defined by the two parameters
$\varphi$ and $\vartheta$ for a given particle number (see
Eq.~\eqref{eq:purestate}).

To get an impression of the purity strength for different initial states the
strongest revival is calculated as a function of $\varphi$ and $\vartheta$.
Since the Bogoliubov backreaction method requires only little numerical effort
we can do this for many initial values.
In this case we use $100$ values for both $\varphi$ and $\vartheta$, thus,
in total $10\,000$ different initial states for each parameter set $\gamma$ and
$g$.
Such a calculation would be out of reach using the quantum jump method to
directly calculate the many-particle dynamics of the master equation.

\begin{figure}
  \includegraphics[width=\columnwidth]{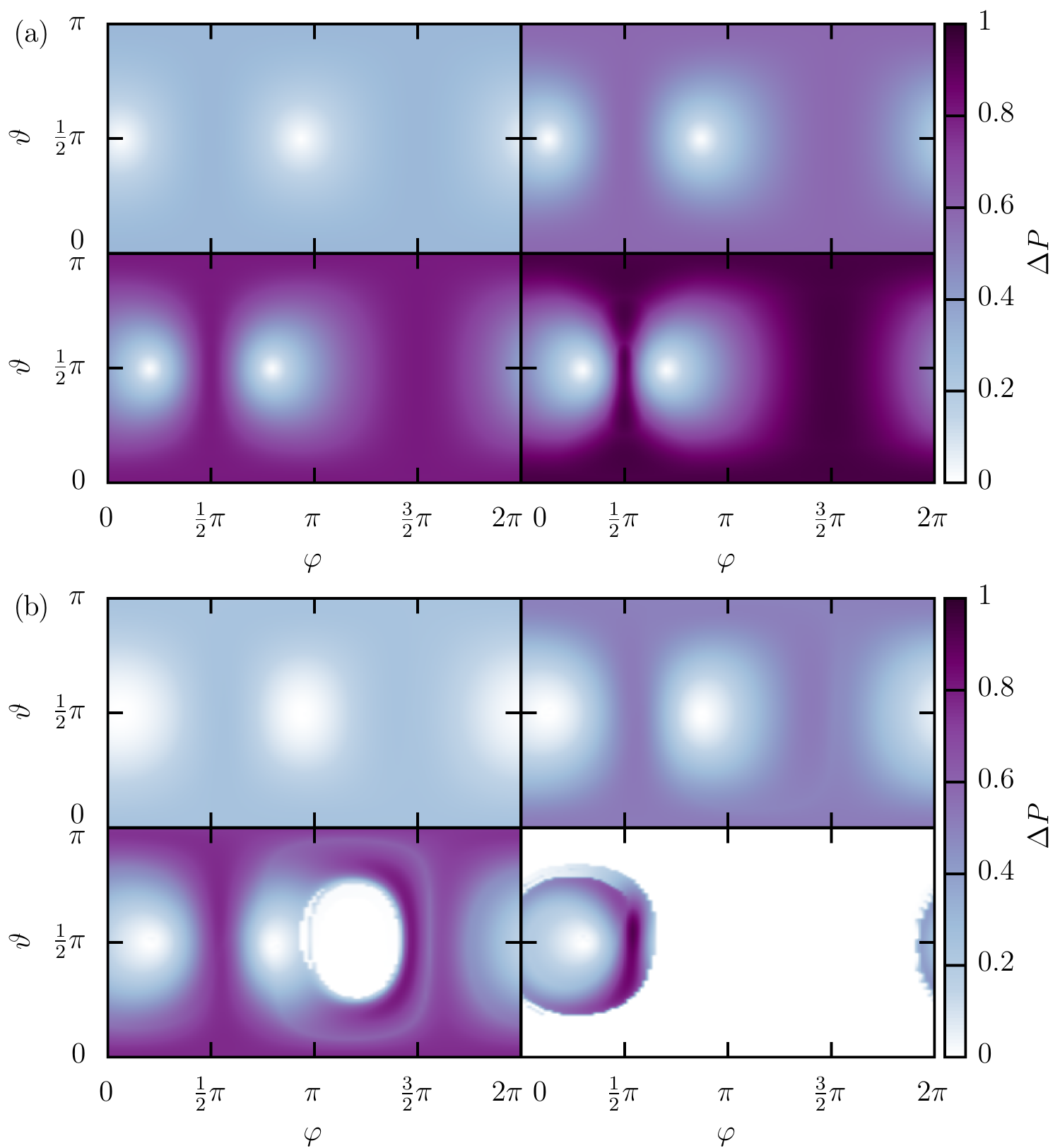}
  \caption{%
    The strongest purity revivals $\Delta P$ for pure initial states defined by
    the two parameters $\varphi$ and $\vartheta$ for (a) $g=0$ and (b) $g=0.5$.
    The gain-loss parameters for both (a) and (b) are $\gamma=0.4$ (upper
    left), $\gamma=0.8$ (upper right), $\gamma=1.2$ (lower left) and
    $\gamma=1.6$ (lower right).
    There are two distinct areas in (a) where the strength of the revivals is
    reduced.
    In the center of these areas lie the stationary ground (left area) and
    excited (right area) states~\eqref{eq:stationaryStates} of the
    $\PT$-symmetric Gross-Pitaevskii equation.
    With interaction (b) an additional region with $\Delta P=0$ arises for
    stronger values of $\gamma$ (lower panels).
    This is the unstable region, in which the particle number diverges and,
    thus, no stable revivals occur.
    For increasing values of $\gamma$ this region expands and for $\gamma=1.6$
    only a small area of initial states with stable revivals survives.
  }
  \label{fig:revivalmap}
\end{figure}

Figure~\ref{fig:revivalmap}(a) shows the strongest revival in the
non-interaction limit $g=0$ for four different values of $\gamma$ which all
show a similar structure.
For most initial states a similar strength of the revivals is found.
As $\gamma$ is increased so is the strength of the revivals.
In the lower right panel the gain-loss parameter is $\gamma=1.6$ and it can be
seen that most initial states even lead to revivals which are close to one.
In these cases the purity is completely destroyed but then is nearly fully
restored.
Furthermore all four panels show two distinct areas where the strength of the
revivals drops to zero.
In the center of these two areas lie the two stationary $\PT$-symmetric states
of the Gross-Pitaevskii equation~\eqref{eq:stationaryStates}.
The left area can be identified with the ground state of the system and the
right area with the excited state.
As discussed in~\cite{Dast16a} these states do not show purity oscillations but
instead the purity decays continuously which explains why the strength of the
revivals vanishes.
In the vicinity of the two stationary states the purity oscillations are less
pronounced and thus the strength of the revivals is also reduced for
nearby states.
It can also be seen that the two areas with weak revivals approach each other
as $\gamma$ is increased.
This is a result of the fact that the two $\PT$-symmetric states coalesce in an
exceptional point at $\gamma=2J$ (here $J=1$).
For $\gamma=2J$ both the ground and excited state are then given by $c_1 =
i/\sqrt{2}$ and $c_2 = 1/\sqrt{2}$ or equivalently $\varphi =
\vartheta = \pi/2$.

The influence of the particle interaction on the purity revivals is shown in
Fig.~\ref{fig:revivalmap}(b).
For small values of the gain-loss parameter $\gamma$ (upper two panels) the
behavior is similar to the non-interaction limit with the difference that the
areas with weak revivals are larger.
In the lower left panel a new type of area arises on the right hand side of the
excited state where no revivals occur.
This is the unstable region where the particle number diverges due to the
particle gain.
One has to be careful when analyzing the revivals in this region since a
diverging state approaches a pure condensate.
This happens since nearly all particles are in the lattice site with gain which
implies $s_z \approx n$ and thus $P\approx 1$.
However, we are only interested in stable revivals and consequently these
unstable revivals are excluded.
For increasing values of the gain-loss parameter $\gamma$ the unstable region
grows and for $\gamma=1.6$ in the lower right panel only a small region of
stable revivals survives in the vicinity of the ground state.
In this parameter region strong oscillations are only found at $\varphi \approx
\pi/2$.

The observation that an unstable region arises close to the excited state which
expands for increasing strength of the gain and loss and finally only a small
stable region near the ground state survives is also found for the
$\PT$-symmetric Gross-Pitaevskii equation and was discussed in detail for a
spatially extended double-well potential in~\cite{Haag14a}.
Note that in this reference an attractive interaction was used in contrast to
the repulsive interaction used in this work, which essentially switches the
roles of the ground and excited state for the stability discussion.

While showing the purity revivals for different initial values gives an
excellent overview, it is less suitable to quantitatively discuss the revival
strength for different parameter values.
To do so we search the initial state that, for constant values of the
parameters $g$, $\gamma$ and $N_0$, leads to the strongest revival,
i.e.\ the maximum in one of the panels of Fig.~\ref{fig:revivalmap}.
This is implemented using a root search which varies $\varphi$ and $\vartheta$
such that
\begin{equation}
  \frac{\partial\Delta P}{\partial\varphi} =
  \frac{\partial\Delta P}{\partial\vartheta} = 0
  \label{eq:rootmaxrev}
\end{equation}
is fulfilled.
Note that $\Delta P$ denotes the maximum revival reached from a specific
initial state which was used in Fig.~\ref{fig:revivalmap} to characterize the
revival strength.
Now we search for the initial state where $\Delta P$ is maximum and name this
quantity $\Delta P_\mrm{max}$.

\begin{figure}
  \includegraphics[width=\columnwidth]{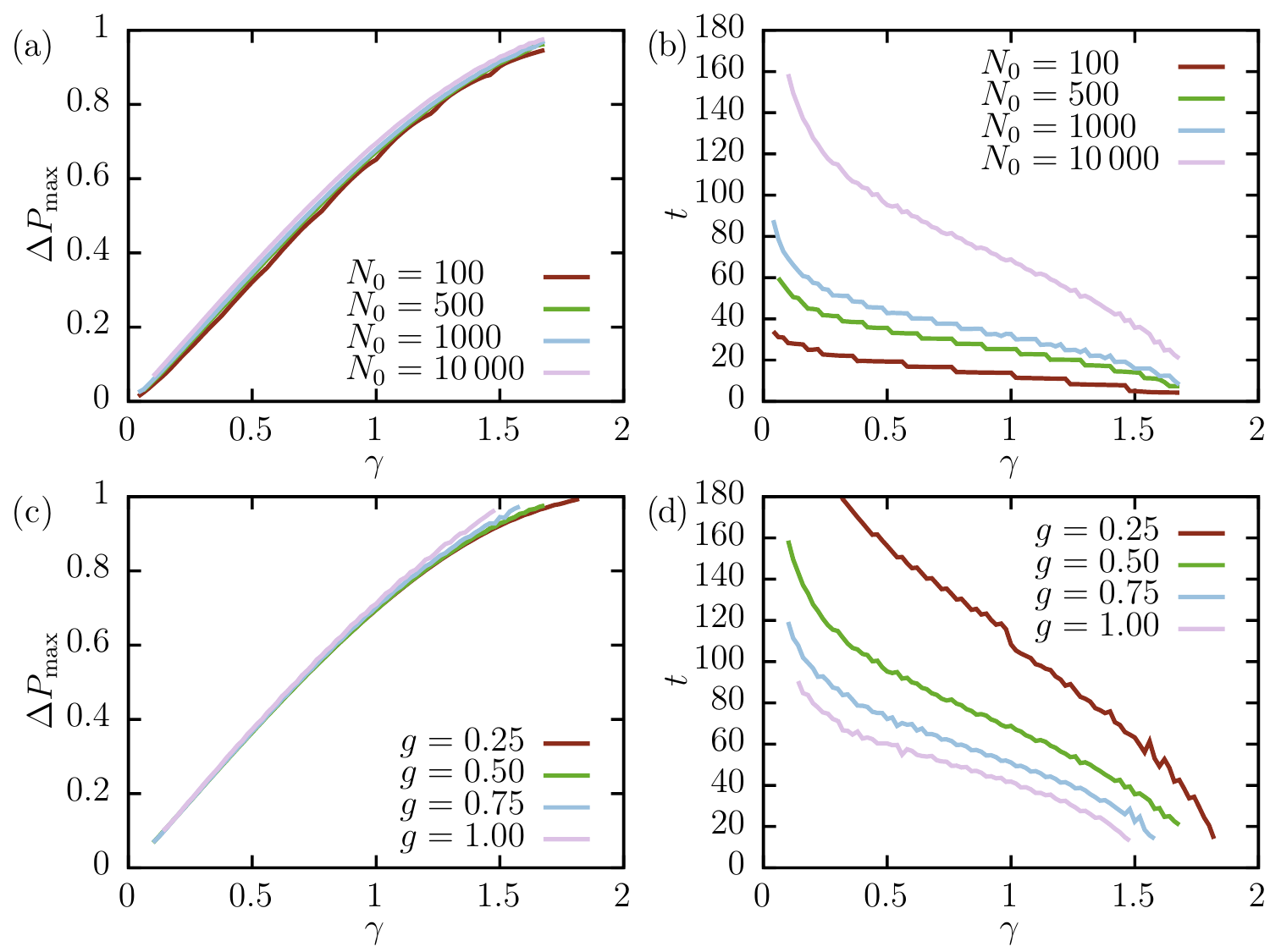}
  \caption{%
    (a) The parameter $\Delta P_\mrm{max}$ is obtained by searching for the
    initial state that leads to the strongest purity revival $\Delta P$.
    The value of $\Delta P_\mrm{max}$ is shown as a function of the gain-loss
    parameter $\gamma$ for four different values of the initial particle number
    $N_0$ and constant interaction strength $g=0.5$.
    The revival strength increases with $\gamma$ but hardly depends on $N_0$.
    (b) The times at which the revivals shown in (a) occur.
    For larger initial particle numbers the strongest revivals occur at later
    times, however, the difference is much smaller compared to the
    non-interacting limit where the time scales linearly with the particle
    number.
    (c) Same as (a) but for a constant large particle number $N_0 = 10\,000$
    and different values of the interaction strength $g$.
    The revival strength is mostly unaffected by the value of $g$.
    (d) The times at which the revivals shown in (c) occur.
    For stronger interaction the maximum revivals occur at smaller times.
  }
  \label{fig:maxrev}
\end{figure}

The maximum value of $\Delta P$ as a function of the gain-loss parameter
$\gamma$ is shown for different initial particle numbers and constant
interaction strength $g=0.5$ in Fig.~\ref{fig:maxrev}(a).
Since the revivals are driven by the gain and loss of the system, the strength
of the revivals increases with the gain-loss parameter $\gamma$.
For $\gamma \to 0$ the revivals vanish and for strong gain and loss, $\Delta
P_\mrm{max}$ is close to one.
However, the remarkable property is that the particle number $N_0$ has almost
no influence on $\Delta P_\mrm{max}$.
This seems counterintuitive at first glance since for $N_0 \to \infty$ the
system can be described by the $\PT$-symmetric Gross-Pitaevskii equation where
the condensate is completely pure and, thus, no purity revivals occur.
However, from the behavior in the non-interacting limit discussed in
Sec.~\ref{sec:linear} we know that different initial particle numbers change
the time scale of the envelope functions of the purity oscillations.
Consequently in this limit the strength of the strongest revival is
approximately the same for different initial particle numbers but only the time
at which these revivals occur changes.

Indeed we also find that with interaction the maximum revivals occur at later
times.
This is shown in Fig.~\ref{fig:maxrev}(b) where the times are shown at which
the maximum revivals $\Delta P_\mrm{max}$ of Fig.~\ref{fig:maxrev}(a) occur.
There is, however, a crucial difference in the scaling behavior of the time.
In the non-interacting limit the time scales linearly with the particle number
and as a result the maximum revivals of $N_0=10\,000$ occur at times that are
larger by a factor of $100$ compared to the revivals of $N_0=100$.
With interaction the difference is much smaller and we find a factor that is
smaller than $10$.

To investigate the influence of the interaction on the purity revivals in more
detail, $\Delta P_\mrm{max}$ is calculated for different values of the
interaction strength $g$ for a constant large particle number $N_0=10\,000$.
Figure~\ref{fig:maxrev}(c) shows that again the actual value of the maximum
purity revivals is mostly unaffected by the interaction strength.
Since neither the particle number $N_0$ nor the interaction strength $g$ has a
significant impact on $\Delta P_\mrm{max}$, we can conclude that the strength
of the maximum revivals is almost entirely determined by the gain-loss
parameter $\gamma$.

The interaction between the particles, however, also has an impact on the times
at which the strongest revivals occur as can be seen in
Fig~\ref{fig:maxrev}(d).
We find that for stronger interactions the maximum revivals occur at shorter
times.
This is an important result since the lifetime of a Bose-Einstein condensate in
an experiment is limited and without the on-site interaction significant purity
oscillations only occur at very large times for a realistic number of
particles.
However, our results show that by adjusting the interaction strength it is
possible to shift these strong purity oscillations towards shorter times.

\section{Eigenvectors of the single-particle density matrix}
\label{sec:eigenvectors}

\begin{figure}
  \includegraphics[width=\columnwidth]{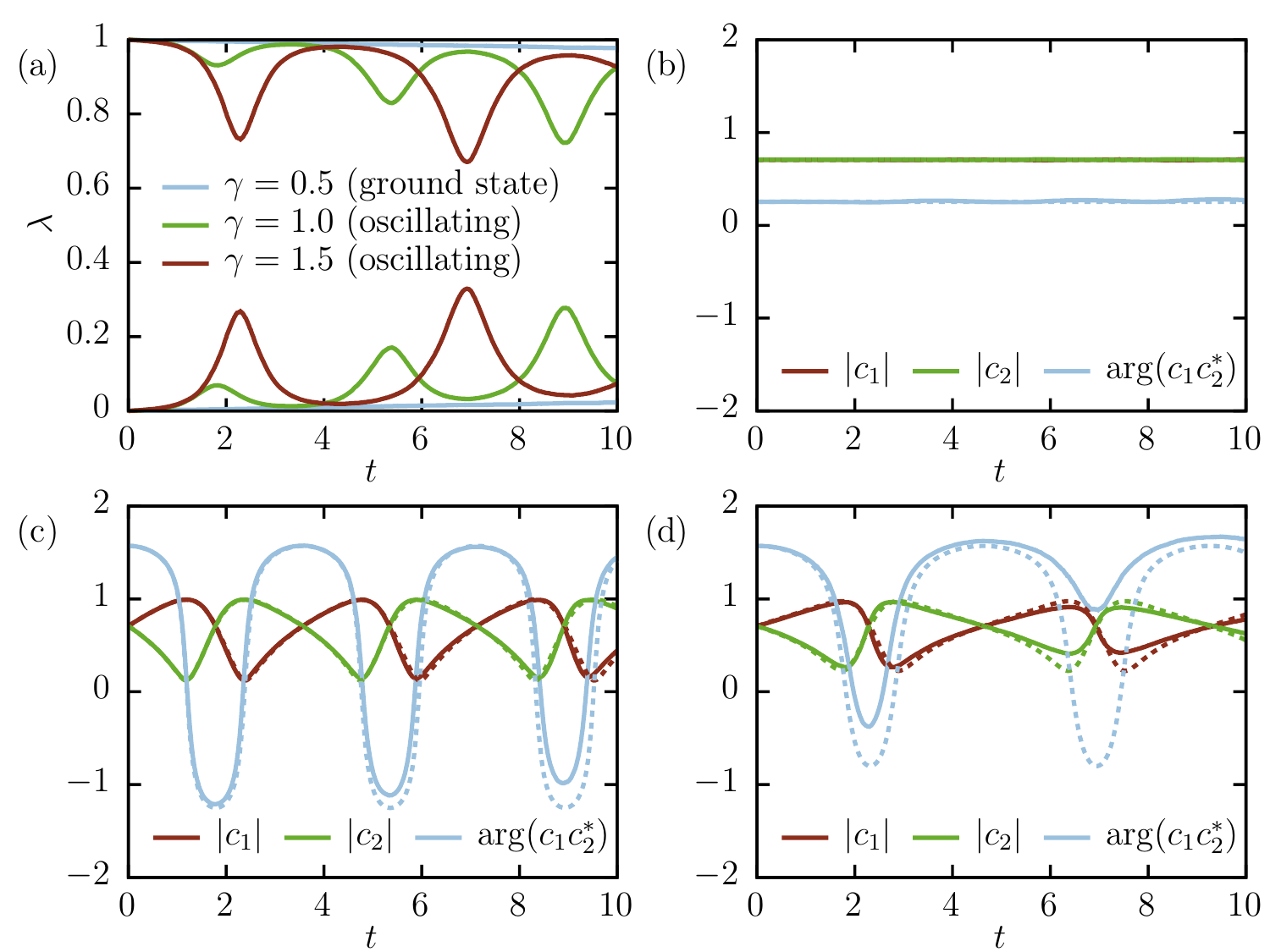}
  \caption{%
    (a) The two eigenvalues of the transposed reduced single-particle density
    matrix using the ground state of the $\PT$-symmetric Gross-Pitaevskii
    equation and the oscillating state $\varphi = \vartheta = \pi/2$ for
    different values of the gain-loss parameter $\gamma$ as initial states.
    The expectation values of the particle number in the two lattice sites
    $|c_j| = \mean{n_j}/N_0$ and the relative phase $\arg(c_1 c_2^*)$ of the
    eigenvector to the greater eigenvalue are shown for (b) the ground state at
    $\gamma=0.5$ and for the oscillating state at (c) $\gamma=1.0$ and (d)
    $\gamma=1.5$.
    The results of the master equation (solid lines) are compared with those
    of the Gross-Pitaevskii equation (dotted lines).
    They show that the eigenvector of the ground state behaves indeed
    stationary when using the master equation (note that $|c_1| \approx
    |c_2|$).
    For the oscillating state the many-particle calculations deviate from the
    mean-field calculations at times where the purity has a dip, and is in good
    agreement at times where the purity is restored.
    In all calculations the parameters $g=0.5$ and $N_0=100$ were used
    and it was averaged over 500 trajectories.
  }
  \label{fig:spdmeigen}
\end{figure}

In this section we study the behavior of the condensed mode.
This can be discussed using the eigenvalues and eigenvectors of the transposed
reduced single-particle density matrix.
The two eigenvalues give the fraction of particles in the condensed and the
non-condensed phase.
For the two-mode system considered here the eigenvalues contain the same
information as the purity and can be written as $\lambda_{1/2} = \frac{1}{2}
(1\pm \sqrt{P})$.
Thus, the eigenvalues show a similar behavior as the purity which can be seen
in Fig.~\ref{fig:spdmeigen}(a) for the stationary ground state and two
oscillating states at different values of the gain-loss parameter.

Since the eigenvector to the macroscopic eigenvalue is the single-particle
state of the condensed phase~\cite{Yang62a} it is possible to directly compare
this eigenvector with the mean-field state of the Gross-Pitaevskii equation.
This extends the discussion in~\cite{Dast14a}, in which expectation values of
the master equation and the $\PT$-symmetric Gross-Pitaevskii equation were
compared.

We will check whether the condensed mode of the many-particle description
behaves stationary if we use the stationary states of the $\PT$-symmetric
Gross-Pitaevskii equation as initial states.
For the ground state this comparison is shown in Fig.~\ref{fig:spdmeigen}(b)
and a similar result can of course be obtained for the excited state.
The comparison shows that indeed also the state itself stays constant.
Note that the underlying density matrix entering the master equation is not
stationary at all but its purity rapidly decays.
Nevertheless the single-particle density matrix stays approximately pure and
the eigenvector to the macroscopic eigenvalue behaves stationary.

Figures \ref{fig:spdmeigen}(c) and \ref{fig:spdmeigen}(d) show the same
comparison for an oscillating state for $\gamma=1$ and $\gamma=1.5$,
respectively.
In the case $\gamma=1$ the normalized expectation values of the particle number
in the $j$'th lattice sites $|c_j| = \mean{n_j}/N_0$ of the many-particle
calculations (solid lines) is in very good agreement with the mean-field
calculations (dotted lines).
The relative phase $\arg(c_1 c_2^*)$, however, shows significant deviations at
precisely the times where the purity of the single-particle density matrix has
a dip.
Consequently the relative phase agrees with the mean-field limit at times where
the purity is restored.
For $\gamma=1.5$ an equivalent behavior is found but the discrepancy to the
mean-field limit is greater especially for the relative phase.

\section{Conclusion}
\label{sec:conclusion}

We have studied a Bose-Einstein condensate with balanced gain and loss
described by a quantum master equation whose mean-field limit is a
$\PT$-symmetric Gross-Pitaevskii equation.
It is already known that the condensate's purity undergoes strong periodic
revivals~\cite{Dast16a}, an effect that cannot be captured using the mean-field
limit where a product state, i.e.\ a pure condensate, is assumed.
These revivals have a direct impact on the contrast measured in interference
experiments.
In this work we deepened the discussion of the purity oscillations by
presenting an analytical solvable model for the non-interacting limit and by
studying the interplay between the interaction strength and the gain and loss
of particles.

To do so we applied the Bogoliubov backreaction method to systems with gain and
loss.
In the non-interacting limit this yields an analytically solvable model for the
elements of the single-particle density matrix.
Since the purity oscillations are driven by the gain and loss of the system
these characteristic oscillations are also found in this limit.
We showed that the frequency of the purity oscillations is mostly unaffected by
the initial amount of particles in the system.
However, the time scale of the envelope functions of the purity has a linear
dependency on the initial particle number.
As a result the time at which strong purity revivals are found is proportional
to the particle number and, thus, is very large for realistic condensates.

With interaction the dynamics described by the Bogoliubov backreaction method
is an approximation and its accuracy has to be checked.
This has been done by a comparison with many-particle calculations using the
master equation and the quantum jump method.
Motivated by this comparison we formulated a limit for the reliability of the
Bogoliubov backreaction method.
Within this limit an excellent agreement between the two approaches was found.

Since the Bogoliubov backreaction method takes into account the backreaction of
the covariances on the single-particle density matrix, it allows us to quantify
the importance of the covariances.
Neglecting this backreaction yields results that differ substantially compared
with the many-particle dynamics.
Thus we can conclude that these covariances are essential to understand the
physics of systems with balanced gain and loss.

The main benefit of the Bogoliubov backreaction method is that it requires only
little numerical effort.
Thus, we were able to characterize the strength of the purity revivals for all
possible initial pure states.
This showed that apart from an area in the vicinity of the $\PT$-symmetric
stationary states of the Gross-Pitaevskii equation strong revivals are found
independent of the initial state.
For stronger gain and loss there is, however, an increasing unstable area where
no stable revivals can be found, and for strong gain-loss contributions only
the area around the $\PT$-symmetric ground state of the mean-field limit is
stable.

Since for the Bogoliubov backreaction method the particle number is only a
parameter that does not change the numerical costs, it is easy to extend the
discussion to larger particle numbers that would not be accessible using the
master equation.
The calculations showed that both the particle number and the strength of the
on-site interaction have little impact on the actual revival strength which
consequently is almost exclusively determined by the strength of the gain and
loss contributions.
As in the non-interacting limit, the times at which the strong revivals occur
increases for larger particle numbers, but with interaction the influence is
considerably smaller.
Increasing the interaction strength even shifts the strong purity oscillations
towards earlier times.
This is an important effect for the experimental observability of the purity
oscillations in systems with balanced gain and loss since the lifetime of a
condensate in an experiment is limited.

Finally the single-particle state of the condensed phase, i.e., the state
corresponding to the macroscopic eigenvalue of the single-particle density
matrix was compared with the mean-field state that enters the Gross-Pitaevskii
equation.
We found a very good agreement between these states at times where the purity
is high, whereas there is a significant discrepancy, especially for the
relative phase, if the purity has a dip.
For the stationary states of the Gross-Pitaevskii equation the single-particle
state of the condensed phase showed a stationary behavior since the
single-particle density matrix is approximately stationary although the
underlying density matrix entering the master equation is not.

As a next step we will study the steady state of the system.
In the non-interacting limit the steady state is an attractor and its form was
given in this work, but it will be interesting to also investigate its behavior
for interacting particles.

\end{document}